\documentclass[12pt]{elsart}
\usepackage{graphics}
\usepackage{epsfig}
\usepackage{setspace}
\usepackage{longtable}

\begin{document}
\begin{frontmatter}
\title{Asymmetric frictional sliding between incommensurate surfaces}
\author{Giuseppe E. Santoro$^{a,b}$\corauthref{cor}},
\ead{santoro@sissa.it} \corauth[cor]{Corresponding Author.}
\author{Andrea Vanossi$^{c}$},
\author{Nicola Manini$^{d}$},
\author{Giorgio Divitini$^{d}$},
\author{and Erio Tosatti$^{a,b}$}
\address{$^a$International School for Advanced Studies (SISSA),
and INFM-CNR Democritos National Simulation Center, Via Beirut 2-4,
I-34014 Trieste, Italy}
\address{$^b$International Centre for
Theoretical Physics (ICTP), P.O.Box 586, I-34014 Trieste, Italy}
\address{$^c$INFM-CNR National Research Center S3, and Department of Physics,
University of Modena and Reggio Emilia, Via Campi 213/A, 41100
Modena, Italy}
\address{$^d$Department of Physics, University of Milan, Via Celoria 16, 20133 Milan, Italy}
\maketitle
\begin{abstract}
We study the frictional sliding of two ideally incommensurate surfaces with 
a third incommensurate sheet -- a sort of extended lubricant -- in between. 
When the mutual ratios of the three periodicities
in this sandwich geometry are 
chosen to be the golden mean $\phi=(1+\sqrt{5})/2$, this system is believed to 
be statically pinned for any choice of system parameters. In the present study
we overcome this pinning and force the two "substrates" to slide
with a mutual velocity $V_{ext}$, analyzing the resulting frictional dynamics. 
An unexpected feature is an asymmetry of the relative sliding velocity of the 
intermediate lubricating sheet relative to the two substrates. 
Strikingly, the velocity asymmetry takes an exactly quantized value which is 
uniquely determined by the incommensurability ratio, and absolutely insensitive
to all other parameters. 
The reason for quantization of the velocity asymmetry will be addressed. 
This behavior is compared and contrasted to the corresponding one obtained for a 
representative cubic irrational, the spiral mean $\omega$.
\end{abstract}

\end{frontmatter}

\section{Introduction}

The intriguing concept of incommensurability has recently found
relevant and practical applications in the discipline of
nanotribology, i.e., the science of friction, lubrication and wear
at the nanoscale, in particular for the possibility of giving rise
to superlubricity \cite{Frenken}. When two hard crystalline
surfaces with incommensurate lattice spacings (or commensurate but
not perfectly aligned) are brought into contact, then the minimal
force required to achieve sliding, known as the static frictional
force $F_s$, vanishes, provided the two substrates are stiff enough. 
This remarkable conclusion can be drawn, for example, in
the context of the Frenkel-Kontorova (FK) model \cite{Aubry1,Aubry2}. 
In real situations however such a case of ``dry'' friction is
exceptional. A physical contact between two solids is generally
mediated by so-called ``third bodies'', which act like a lubricant
film. The sliding interface is thus characterized by three
distinct length scales: the periods of the bottom and top
substrates, and the period of the lubricant structure. 
Upon relative sliding of the two substrates, physical intuition would
suggest that this confined lubricant should move with the average
of the two substrate velocities. At the nanoscopic level, however,
incommensuration between the atomic corrugations of the solid
substrates could change this expectation, with possible
tribologically important implications of an irregular distribution
of the lubricant velocity on the wear of the sliding surfaces.

\begin{figure}[p]
\centerline{
\epsfig{file=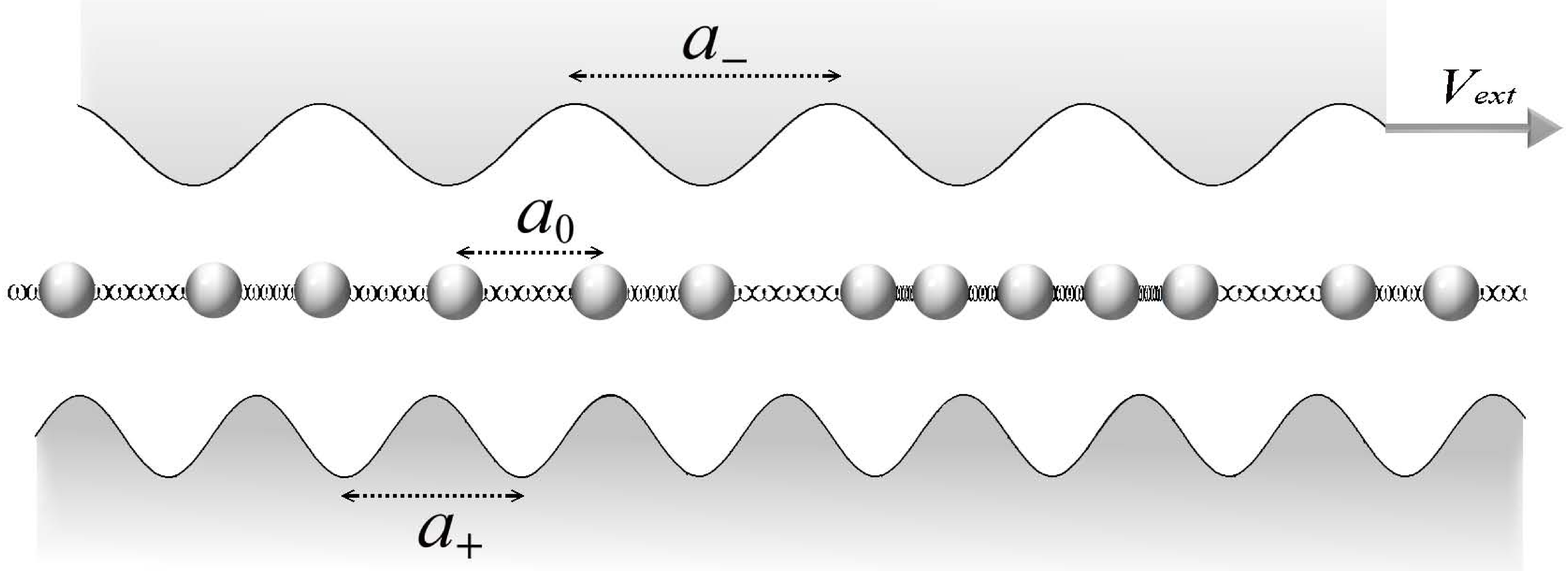,width=15cm,angle=0} 
}
\caption{\label{fig-model} Schematic drawing of the model with three characteristic
length scales. Here $a_0$ is to be interpreted as the equilibrium distance of the
harmonic chain (lubricant film). $a_{\pm}$ are the spatial periods of the two
sinusoidal substrate potentials, assumed to be rigid.} 
\end{figure}

In order to study the role of incommensurability among the three
interface inherent lengths on the sliding dynamics, we consider, as
in Ref.~\cite{Braun}, a simplified one-dimensional generalized FK model 
consisting of two moving rigid sinusoidal substrates, of spatial periodicity
$a_{+}$ and $a_{-}$, and a chain of harmonically interacting particles, 
of equilibrium length $a_0$, mimicking the lubricant and sandwiched between the 
two substrates, as shown in Fig.~\ref{fig-model}. 

In the standard FK model (single substrate potential, of period $a_+$, plus harmonic chain), 
the static behavior depends strongly on the value of $r=a_+/a_0$: a rational $r$ produces 
pinning (a finite force is necessary to induce lubricant translation), while an irrational 
$r$ is associated to a phase transition (the Aubry transition) between a pinned phase, 
for small spring stiffness $K$, to an unpinned regime, for large $K$ \cite{Aubry83}.
A clear critical behavior was demonstrated only for quadratic algebraic irrational numbers, 
among which the Aubry critical $K_c$ for depinning is smallest for $r=\phi\equiv(1+\sqrt{5})/2$,
the golden mean (GM) \cite{Greene78}.
In this sense, therefore, the GM represents the ``most irrational'' incommensuration, with the 
largest unpinned phase.
 
We will show that, on the contrary, the dynamics of our generalized FK model shows
a surprising result, the Golden Mean case leading to a strong {\em dynamical} 
commensuration.  

\section{Model}

Consider the system sketched in Fig.~\ref{fig-model}. 
The equation of motion of the $i$-th lubricant particle is:
\begin{eqnarray} \label{eqmotion:eqn}
m\ddot{x}_i &\,=\,& -\frac{1}{2} \left[ F_+ \sin{k_+ (x_i-v_+t)} + F_- \sin{k_-(x_i-v_-t)}\right] 
\nonumber \\
&& \hspace{1mm} + \, K (x_{i+1}+x_{i-1}-2x_i) - \, 2\gamma (\dot{x}_i - v_a) \;,
\end{eqnarray}
where $m$ is its mass, $K$ is the chain spring constant,
$k_{\pm}=(2\pi)/a_{\pm}$ are the wave-vector periodicities
associated to the two sinusoidal substrates of amplitude $U_{\pm}$, 
moving at velocities $v_{\pm}$. 
Here $v_a=(v_+ + v_-)/2$ is the average speed of the substrates, 
and $\gamma$ is a viscous friction constant accounting phenomenologically for 
various sources of dissipation in the substrates (electrons, phonons, etc.). 
(The last term in Eq.~\ref{eqmotion:eqn} originates from two frictional contributions of the
form $-\gamma(\dot{x}_i - v_+) - \gamma(\dot{x}_i - v_-)$.) 
$F_{\pm}=k_{\pm}U_{\pm}$ are the amplitudes of the forces due to the corrugation of the 
substrates, and we set $F_-/F_+=1$ as the least biased choice. 
The spring equilibrium length $a_0$ (not entering explicitly the 
equation of motion (\ref{eqmotion:eqn})) enters through the boundary
conditions, which we take to be periodic (PBC), $x_{N_0+1}=x_1+N_0\, a_0$.
The traditional implementation of the PBC \cite{Braun,Vanossi} is realized
by approximating $a_-/a_+$ and $a_+/a_0$ 
(via, for instance, a continued fraction expansion \cite{Khinchin})
with suitable rational numbers $N_-/N_+$ and $N_+/N_0$ numerically close to $r$ 
(e.g.\ successive integers $N_-$, $N_+$, $N_0$ in the Fibonacci sequence approximate $r=\phi$).
Convergence of all physical quantities is then checked in the limit of
large $N_{\pm}$ and $N_0$, where full incommensurability is restored.
In the present case, however, the boundary condition enters only in the $K$-term, 
and there is no real need to approximate $r$: we can use the machine-precision value 
of $r$. 
This alternative procedure introduces, at the boundary, a small phase ``twist''
to the purely local substrate forces, but this boundary error can be made arbitrarily 
small by choosing a suitably large number $N_0$ of particles.
All the results presented below coincide for either implementation of the PBC.

The general behavior of the driven system in Eq.~(\ref{eqmotion:eqn}) depends 
crucially on the nature of the incommensurability of the substrates and the chain. 
In order to make a comparison with previous tribological studies
\cite{Braun,Vanossi}, as well as to previous studies on generalized FK 
models \cite{Fasolino}, we restrict our present analysis to cases
where the ratio $a_-/a_+$ of the substrate periodicities and the
ratio $a_+/a_0$, fixing the equilibrium length of the lubricant
chain, are equal: $r=a_-/a_+=a_+/a_0>1$. 
We take $a_+=1$, $m=1$, and $F_+=1$ as basic dimensionless units. 

The equations of motion are integrated using a standard fourth-order 
Runge-Kutta algorithm. The system is initialized with the chain particles
placed at rest at uniform separation $a_0$. After relaxing the
starting configuration and selecting a reference frame in which
the $U_+$ substrate is at rest ($v_+=0$), the top substrate $U_-$ begins to
move at the constant velocity $v_-=V_{\rm ext}$. 
After an initial transient, the system reaches a ``dynamical stationary state'',
characterized by generally fluctuating but significantly (on
average) constant drift velocity of all particles.  

Deferring a more general discussion to later work, we address here
the contrasting behavior of two different incommensurabilities $r$, namely, 
the golden mean (GM) and the spiral mean (SM)
\footnote{The golden mean $\phi\equiv(\sqrt{5}+1)/2\approx 1.6180$
is the irrational number satisfying the algebraic quadratic
equation $\phi^2-\phi-1=0$; its rational approximants
\cite{Khinchin} are provided by the Fibonacci sequence,
$F_{n+1}=F_{n}+F_{n-1}$, with $F_0=F_1=1$: $F_{n+1}/F_n\to \phi$. 
The spiral mean ($\omega\approx 1.3247$) belongs to the class of cubic irrationals,
satisfying the equation $\omega^3-\omega-1=0$; 
its rational approximants can be generated by the recursion relation
$G_{n+1}=G_{n-1}+G_{n-2}$ with $G_{-2}=G_0=1,G_{-1}=0$: $G_{n+1}/G_n\to \omega$},
belonging to distinct classes of irrational numbers (quadratic and cubic, respectively). 
The very different behavior of golden and spiral will demonstrate how strongly 
the precise value of incommensurability dramatically influences the tribological
behavior of interface sliding.

Statically \cite{Fasolino,Vanossi}, 
it was found that when the length scale ratio $r$ equals the SM there
exists a value of the harmonic interaction strength $K$,
above which the static friction force $F_s$ is zero. 
On the contrary, when $r$ is equal to the GM, then $F_s$ is nonzero for
any value of $K$: the system is always pinned. 

Dynamically, the same model was recently considered with the two
substrates forced to slide relative to one another by an external
force applied through an elastic spring \cite{Braun}. It was found
that the GM and the SM behaved very differently, showing that
there is no quantitative or even qualitative uniformity of
behavior in incommensurate kinetic friction, and that certain
incommensurabilities favor sliding systematically better than others.

In the present study of dynamic sliding we focus on the motion of the
lubricant film and analyze the corresponding particle motion. 
The sliding of one substrate relative to the other is rigidly forced, in the form of
a fixed relative velocity (no elastic pulling, as in Ref.~\cite{Braun}).

\section{Results and discussion} \label{results:sec}

\begin{figure}
\centerline{
\epsfig{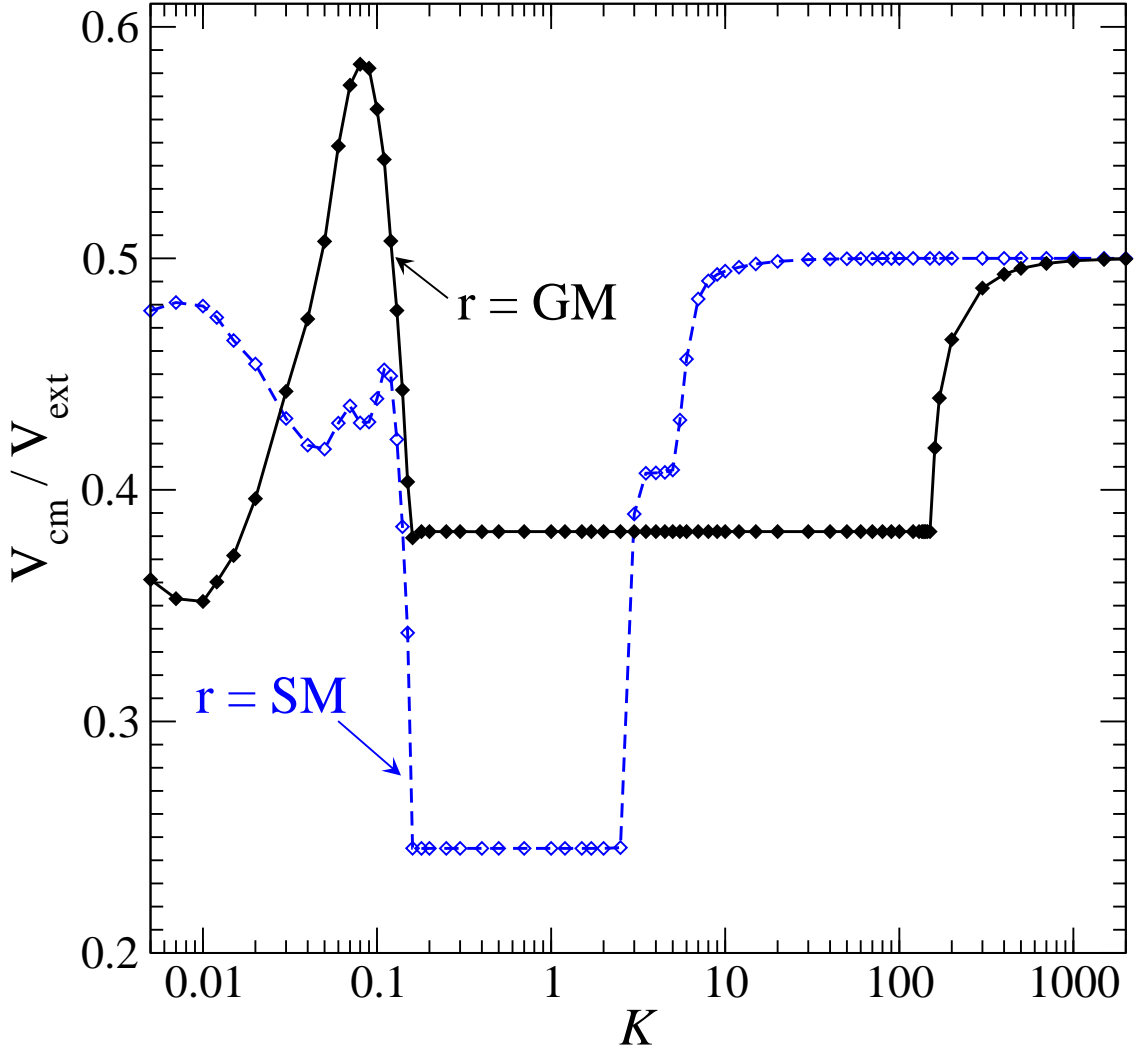} 
}
\caption{\label{fig-velcm_GS} Normalized velocity of the center of mass,
$V_{cm}/V_{ext}$, as a function of the chain stiffness $K$,
for golden mean (GM) and spiral mean (SM) incommensurability. 
Here $\gamma=0.1$ and $V_{ext}=0.1$. Note the logarithmic scale in the abscissa.
Observe that $V_{cm}/V_{ext}$ approaches the expected value $1/2$ only for large 
values of $K$ (hard lubricant). 
The value of the wide GM plateau is exactly tuned by a dynamical matching condition
which results in $V_{cm}/V_{ext}=1/(1+\phi)\approx 0.381966$ 
(see Eq.~\protect\ref{v:eqn}).}
\end{figure}

Figure~\ref{fig-velcm_GS} shows the time-averaged lubricant chain center of mass (CM) 
velocity $V_{cm}$, after stationarity has been reached, as a function of the chain
stiffness $K$ for the GM and SM values of $r$. 
The value of $\gamma$ is here fixed to be $\gamma=0.1$ (under-damped regime), and 
the external velocity is $V_{ext}=0.1$.
The first obvious observation is that only in the limit of large
$K$ does the lubricant move at the mean speed $v_a=V_{ext}/2$:
generally, the CM velocity $V_{cm}$ is a complicated function of $K$. 
It encompasses regimes of continuous evolution, but also remarkably and 
unexpectedly flat plateaus.

\begin{figure}
\centerline{
\epsfig{file=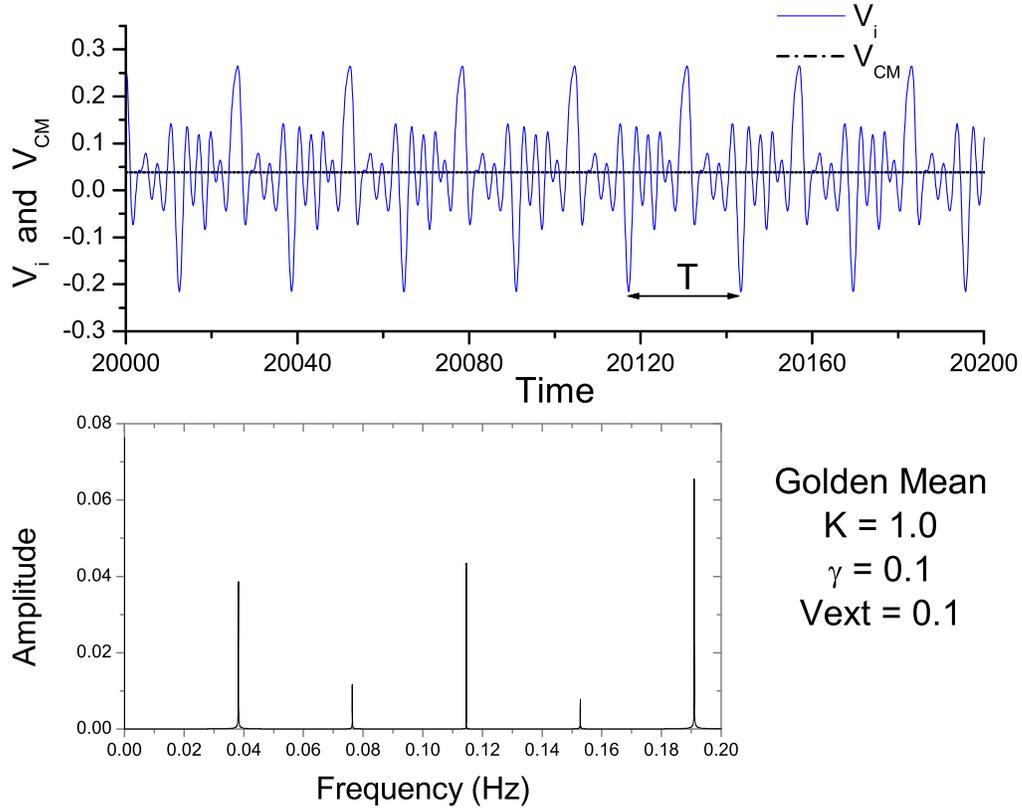,width=15cm,angle=0} 
}
\caption{\label{fig-Golden} Typical time-evolution of the velocity of a particle
$V_i$, and of the center of mass $V_{cm}$, for a value of $K$ inside the GM 
plateau ($K=1$).  
$T$ indicates the time periodicity of the sliding dynamics at stationarity, 
exactly given by $T = a_+ (1+\phi)/V_{ext} \approx 26.180$ 
(see Sec.~\protect\ref{results:sec}).
The lower panel displays the extraordinarily sharp peaks of the
corresponding Fourier transform of $V_i$, showing a single frequency, exactly located 
(see Sec.~\protect\ref{results:sec})
at $f_+=T^{-1}=V_{ext}a_+^{-1}/(1+\phi)\approx 0.0381966$ 
(plus higher harmonics).} 
\end{figure}

\begin{figure}
\centerline{
\epsfig{file=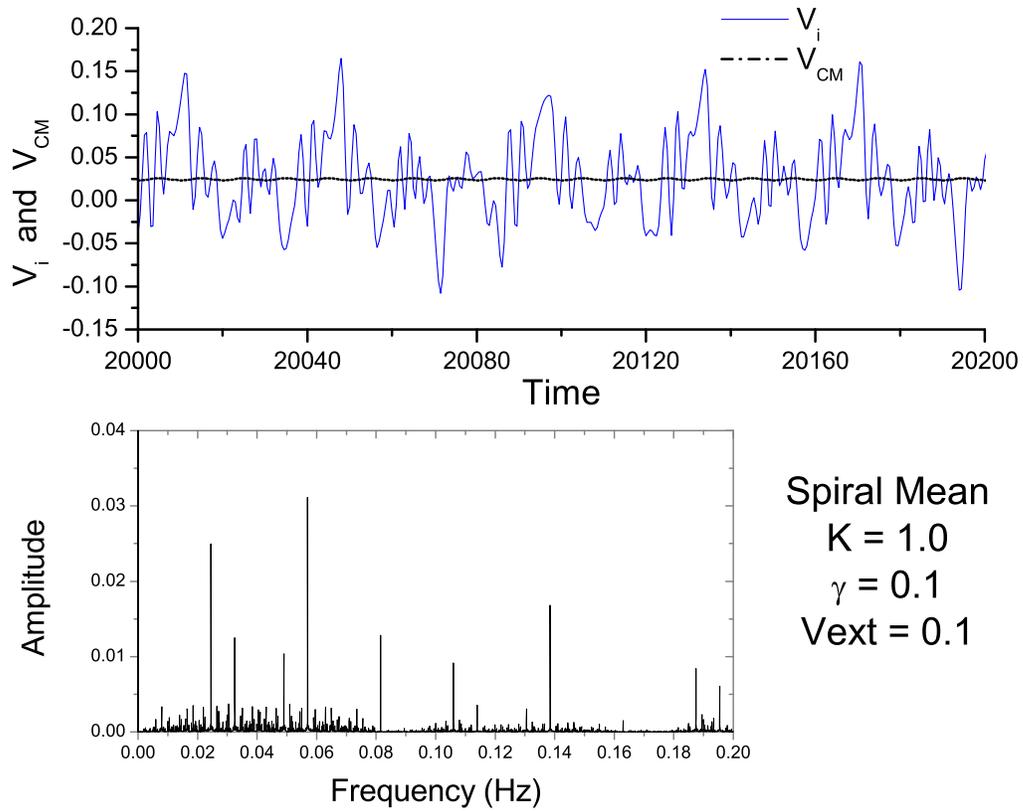,width=15cm,angle=0} 
}
\caption{\label{fig-Spiral} Same as Fig.~\ref{fig-Golden}, but for the
SM incommensurability, still for a value of $K$ inside a plateau ($K=1$). 
At variance with Fig.~\ref{fig-Golden}, no periodic dynamics is here observed, 
as the noisy corresponding Fourier spectrum (lower panel) gives
further evidence of.}  
\end{figure}

The most striking feature is indeed the presence of perfectly flat $V_{cm}/V_{ext}$ plateaus, 
whose precise value is remarkably independent not only of $K$, but also of $\gamma$,
$V_{ext}$, and even (we checked) of $F_-/F_+$. 
Superficially, the two incommensurability considered, GM and SM, appear to share the 
same general features.
Going deeper into the details of the particle motion, the two cases appear to be somewhat
different.  
In order to clarify these differences, we plot in Figs.~\ref{fig-Golden} and \ref{fig-Spiral} 
the typical time-evolution of the velocity of a particle and of $V_{cm}$, for a value of 
$K$ inside a plateau: $K=1$. 
For the GM case ($r=\phi$), the motion of each single particle is perfectly time-periodic 
with period $V_{ext}\, T \simeq 2.62$ (see Fig.~\ref{fig-Golden}, in particular the
Fourier spectrum panel).
Like $V_{cm}/V_{ext}$, also $V_{ext}\, T$ is found to be independent of all parameters.
The motion of every particle in the chain is simply characterized by 
a phase-variable $\theta_i$, 
so that we can write, in the stationary state, $v_i(t) = g(\omega t -\theta_i)$ where
$g(z)$ is a periodic function in $[0,2\pi]$, and $\omega=2\pi/T$.
On the contrary, the motion is not periodic in the SM case, as clear from the Fourier
spectrum panel in Fig.~\ref{fig-Spiral}.

The key to the exact quantization, periodicity of motion, and universality of 
$V_{cm}/V_{ext}$ and $V_{ext}\, T$ for the GM case, is what we might call a form 
of {\em dynamical commensurate sliding}, which, surprisingly, the GM incommensurability
shares with all the rational values of $r$. 
We now illustrate this effect. 
Consider traveling with a given center of mass velocity $0<V_{cm}<V_{ext}$. 
Correspondingly, let $f_+=T_+^{-1}=V_{cm}/a_+$ and $f_-=T_-^{-1}=(V_{ext}-V_{cm})/a_-$ be 
the average frequency of encounter with the substrate corrugation minima.
It turns out that, for the GM case, the $V_{cm}$ is perfectly tuned so as to make
the two frequencies $f_+$ and $f_-$ mutually commensurate, indeed perfectly equal:
\begin{equation} \label{matching:eqn}
f_+=f_- \qquad \hspace{10mm} \mbox{(Golden Mean matching)} \;,
\end{equation}
so that the particles can ``satisfy'' both substrates by a concerted motion with period 
$T=f_+^{-1}=a_+/V_{cm}$. 
It is a matter of simple algebra to realize that Eq.~\ref{matching:eqn} is fulfilled
when the average lubricant speed $V_{cm}$ is given by:
\begin{equation} \label{v:eqn}
\frac{V_{cm}}{V_{ext}} = \frac{a_+}{a_+ + a_-} = \frac{1}{1+\phi} \;.
\end{equation}
Correspondingly, the period of the motion is given by $V_{ext} T = a_+ (1+\phi)$,
and the fundamental frequency of the Fourier spectrum by $f_+=V_{ext}a_+^{-1}/(1+\phi)$.
Eq.~\ref{v:eqn} describes precisely the value of the plateau in $V_{cm}/V_{ext}$ 
show in Fig.~\ref{fig-velcm_GS} for the GM case. 
Similarly, the value of $f_+$ is precisely the fundamental frequency shown in the Fourier 
panel of Fig.~\ref{fig-Golden}.
The GM plateau in $V_{cm}/V_{ext}$ is remarkably robust and wide (indeed, its
width as a function of $K$ increases for decreasing $V_{ext}$): we have found no
other irrational with a similarly wide plateau, which definitely competes with
those obtained for rational values of $r$ (not shown). 
In the present three-slider dynamical context, therefore, the GM ratio 
represents the most ``{\em dynamically commensurate} irrational''. 
That is just the opposite of the statics of the standard FK model \cite{Aubry1,Aubry2}
where GM incommensurability is the hardest one to pin.

The SM case is quite different. The plateau is not just smaller:
it is not quantized by any frequency matching condition as in Eq.~(\ref{matching:eqn}).
Correspondingly, the particle motion in the chain is definitely not periodic,
as confirmed by the Fourier spectrum panel in Fig.~\ref{fig-Spiral}.

\section{Conclusions}

When two periodic substrates are forced to slide relative to one
another but there is an interposed ``lubricant'' layer in between,
the velocity of the intermediate layer may display an interesting
behavior. In general, this velocity will not be the average between
that of the two substrates. Moreover, in some cases at least, the
lubricant layer velocity may be quantized to an exact value,
depending only on relative commensurabilities. 
That occurs not only if the two substrates and the layer are mutually
commensurate (rational values of $r=a_-/a_+=a_+/a_0$),  but strikingly also 
when they have certain types of incommensurabilities, for instance their periods 
are related through the golden mean, $r=\phi$. 
On the other hand, that certainly does not apply to all types of incommensurability, 
and we found no simple matching quantization and a radically different behavior (aperiodic
motion) in the case of the spiral mean, $r=\omega$. 
Work is under way to fully explore this interesting scenario.

\section*{Acknowledgments}
We are grateful to O.M Braun for collaboration, stimulation, and illuminating discussions. 
This work was partly supported by MIUR COFIN No.\ 2003028141, MIUR COFIN No.\ 2004023199, 
by  FIRB RBAU017S8R operated by INFM-CNR, by MIUR FIRB RBAU017S8 R004, and by
INFM-CNR (Iniziativa trasversale calcolo parallelo). 
A.V.\ was supported in part by PRRIITT (Regione Emilia Romagna), Net-Lab
``Surface \& Coatings for Advanced Mechanics and Nanomechanics'' (SUP\&RMAN).


\end{document}